\documentclass[twocolumn,superscriptaddress,aps,showpacs,floatfix,prl,10pt]{revtex4-1}
\usepackage[T1]{fontenc}                                                                                                                                                                                                   
\usepackage{amssymb,amsmath}
\usepackage{graphicx}
\usepackage{color}
\usepackage{bbold}
\usepackage{soul}
\usepackage[dvipsnames]{xcolor}
\usepackage{ulem}
\newcommand{\bs}[1]{{\boldsymbol{#1}}}
\newcommand{\bk}{\bs{k}}
\newcommand{\br}{\bs{r}}
\newcommand{\av}[1]{\overline{#1}}
\newcommand{\tauh}{\tau_H}

\newcommand{\ccfs}{C_\mathrm{F}}
\newcommand{\ccbs}{C_\mathrm{B}}
\newcommand{\ket}[1]{\left| #1 \right\rangle}

\def\rmi{\mathrm{i}}
\def\rme{\mathrm{e}}

\begin{document}
\title{ Coherent forward scattering as a signature of Anderson metal-insulator transitions}
\author{Sanjib Ghosh} 
\affiliation{Centre for Quantum Technologies, National University of Singapore, 3 Science Drive 2, Singapore 117543, Singapore}
\affiliation{MajuLab, UMI 3654, CNRS-UNS-NUS-NTU International Joint Research Unit, Singapore}
\affiliation{Laboratoire Kastler Brossel, UPMC-Sorbonne Universit\'es, CNRS, ENS-PSL Research University, Coll\`{e}ge de France, 4 Place Jussieu, 75005 Paris, France}
\author{Christian Miniatura}
\affiliation{MajuLab, UMI 3654, CNRS-UNS-NUS-NTU International Joint Research Unit, Singapore}
\affiliation{Centre for Quantum Technologies, National University of Singapore, 3 Science Drive 2, Singapore 117543, Singapore}
\affiliation{Department of Physics, National University of Singapore, 2 Science Drive 3, Singapore 117542, Singapore}
\affiliation{School of Physical and Mathematical Sciences, Nanyang Technological University, Singapore 637371, Singapore}
\affiliation{Universit\'{e} C\^ote d'Azur, CNRS, InPhyNi, Valbonne F-06560, France}
\author{Nicolas Cherroret}
\affiliation{Laboratoire Kastler Brossel, UPMC-Sorbonne Universit\'es, CNRS, ENS-PSL Research University, Coll\`{e}ge de France, 4 Place Jussieu, 75005 Paris, France}
\author{Dominique Delande}
\email{e-mail: Dominique.Delande@lkb.upmc.fr}
\affiliation{Laboratoire Kastler Brossel, UPMC-Sorbonne Universit\'es, CNRS, ENS-PSL Research University, Coll\`{e}ge de France, 4 Place Jussieu, 75005 Paris, France}

\begin{abstract}
We show that the coherent forward scattering (CFS) interference peak amplitude sharply jumps from zero to a finite value upon crossing a metal-insulator transition. Extensive numerical simulations reveal that the CFS peak contrast obeys the one-parameter scaling hypothesis and gives access to the critical exponents of the transition. We also discover that the critical CFS peak directly controls the spectral compressibility at the transition where eigenfunctions are multifractal, and we demonstrate the universality of this property with respect to various types of disorder.
\end{abstract}

\pacs{05.60.Gg, 72.15.Rn, 42.25.Dd, 03.75.-b}

\maketitle

About sixty years ago, P. W. Anderson established that interference can completely suppress diffusion \cite{Anderson58}. Later, it was even predicted that three-dimensional (3D) systems exhibit a genuine disorder-driven metal-insulator transition (MIT) \cite{Edwards72, Abrahams79}. Since then, various classes of MITs, with different critical properties, have been  identified \cite{Evers08, Dobrosavljevic12}. Generically, a MIT features a mobility edge separating a metallic phase, where waves are extended and propagate diffusively, from an insulating phase where waves are localized. Recently observed in spinless time-reversal invariant systems~\cite{Hu08, Aubry14, Chabe08, Jendrzejewski3D12, Semeghini15}, Anderson MITs still remain challenging and elusive in more exotic configurations where time-reversal or spin-rotation is broken, or when interactions are present \cite{Schreiber15, Choi16}. Furthermore, transport properties near the critical point, affected by the multifractal character of the eigenstates~\cite{Mirlin10}, have been little studied in actual experiments~\cite{Faez09}.

\begin{figure}[h!]
 \includegraphics[width=0.92\columnwidth]{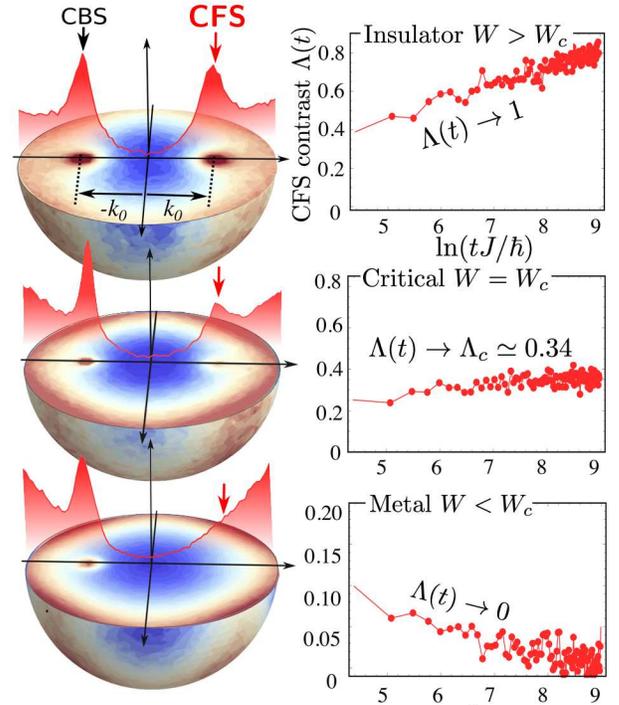}
 \caption{(Color online)
Left: long-time limit of the disorder-averaged momentum distribution $n(\bk,t)$ obtained in the insulating (top, $W=20J$), critical (middle, $W=16.5J$) and metallic (bottom, $W=12J$) phases of the cubic 3D Anderson model (lattice constant $a$, tunneling rate $J$) when an initial plane wave $\ket{\bk_0}$ is numerically propagated up to time $t=8000\hbar/J$. The energy is set to $E=J$ and $\bk_0=\pi/(3a) \hat{\bs{e}}_x$. The CFS and CBS peaks are visible at $\bk_0$ and $-\bk_0$ respectively. The solid curve is a cut along $k_x$.  
Right: time evolution of the normalized CFS contrast $\Lambda$ in the three phases.}
 \label{3DMomentumDistribution}
\end{figure}

Related to Anderson localization (AL), coherent forward scattering (CFS) is a robust interference effect which triggers a macroscopic peak in the forward direction of the momentum distribution $n(\bk,t)$ obtained after an initial plane wave $\ket{\bk_0}$ has evolved through a bulk disordered system \cite{Karpiuk12, Lee14, Ghosh14}. While CFS resembles the well-known coherent backscattering (CBS) effect, which is due to the pair interference of time-reversed scattering sequences and yields a peak in the backward direction \cite{Cherroret12}, the two effects turn out to be fundamentally different. Indeed, the CBS peak relies on time-reversal symmetry (TRS) \cite{Tiggelen98} and exists on both sides of the MIT, with no discontinuous behavior as the mobility edge is crossed~\cite{Ghosh15}. In marked contrast, CFS \textit{requires} Anderson localization to show up (it is absent in the metallic phase) and is present whether or not TRS is broken~\cite{Micklitz14, Lemarie16}. While the experimental observation of CBS in momentum space has been recently achieved with cold atoms \cite{Jendrzejewski12}, no observation of CFS has been reported so far. On the theoretical side, CFS has been  studied in one dimension and two dimensions \cite{Karpiuk12, Ghosh14, Micklitz14}, but not in three dimensions where an Anderson MIT takes place. In this article, we numerically demonstrate that the CFS contrast constitutes a reliable and measurable order parameter for MITs: (i) it jumps abruptly from zero in the metallic phase to a finite value in the insulating phase, (ii) obeys the one-parameter scaling hypothesis and gives access to the critical exponents of the transition, and (iii) directly controls the spectral compressibility at the transition, where eigenfunctions are multifractal. Using large-scale numerical investigations, we prove that the latter property is universal and we validate the conjecture that links the critical spectral compressibility to the fractal information dimension of the Anderson MIT in the orthogonal Gaussian Ensemble (GOE).

In the left panel of Fig. \ref{3DMomentumDistribution}, we show a density plot of the momentum distribution $n(\bk,t)$ resulting from the numerical propagation over long times of an initial plane wave $\ket{\bk_0}$ in a 3D random potential, in the insulating (top), critical (middle) and metallic phases (bottom). While the CBS peak at $\bk=-\bk_0$ is present in the three phases, the CFS peak at $\bk=\bk_0$ only exists in the critical and insulating regimes.
These results have been obtained with the 3D tight-binding Anderson Hamiltonian $\hat{H}=-J \sum_{\langle i j \rangle}c_i^\dagger c_j+\sum_i V_i c_i^\dagger c_i$, with nearest-neighbor hopping only (strength $J$) and periodic boundary conditions. 
Lattice sites $i$ and $j$ run over a simple 3D cubic lattice comprising $M^3=120^3$ sites with spacing $a$. The system is virtually infinite as its size is larger than the longest distance traveled during the simulations. The random onsite potential energies $V_i$ are taken from the distribution  $P(V_i)=1/W$ within $[-W/2, W/2]$, and correlations between sites $i$ and $j$ are described by the correlation function $C_{ij}=\overline{V_iV_j}=\overline{V_i^2}\delta_{ij}$. 
We select a given energy $E$ by applying the filtering operator $F_\sigma(E)=\exp[-(E-\hat{H})^2/2\sigma^2]$ onto the initial plane wave state $\ket{\bk_0}$. 
Following the most accurate known numerical results~\cite{Slevin14}, we choose $E=J$ and vary the disorder strength around the critical point $W_c(E)\approx 16.53J$ of the Anderson MIT. The width of the filter is $\sigma=0.5J$, so that $W_c$ is almost independent of $E$ in the selected range. We then evolve this filtered state with $\mathcal{U}(t) = \exp({-i\hat{H}t/\hbar}).$ Our numerical scheme uses an expansion of the filtering and evolution operators in terms of Chebyshev polynomials of $\hat{H}$, see \cite{Ghosh14} for details. This procedure is repeated for 6000 different disorder configurations to compute the disorder-average momentum distribution $n(\bk,t)=\av{|\langle{\bk}|\mathcal{U}(t)F_\sigma(E)\ket{\bk_0}|^2}$. 

Let us now discuss the time dynamics of CBS and CFS across the MIT. The evolution of the CBS peak is simple: whatever $W$, this peak becomes sizable after a few mean free times $\tau$ \cite{Ghosh15} and its amplitude shows \textit{no discontinuity} as the mobility edge is crossed. To analyze the dynamics of CFS, we use the CFS and CBS contrasts $\ccfs$ and $\ccbs$ (defined as the peak height above the background of the momentum distribution at $\pm \bk_0$ over the background) \cite{Ghosh14}. 
As shown below, the CFS contrast $\ccfs$ is a smoking gun of Anderson localization, as it vanishes at long times in the localized regime and grows to a large value -- of the order of unity -- in the localized regime. The CBS contrast
behaves very differently, as it is not singular at the Anderson transition: it is almost exactly unity in the
diffusive regime and slowly decreases far in the localized regime. We thus chose to compute the normalized contrast $\Lambda(W,t) = \ccfs/\ccbs$ (between 0 and 1). This definition proves less sensitive to statistical fluctuations of the background than $\ccfs$ and $\ccbs$ themselves. The same conclusion and similar quantitative results, although a bit more noisy, are obtained if one uses
$\ccfs$ only as the critical quantity. In a system where time-reversal symmetry is broken~\cite{Lemarie16}, the CFS peak is still present, but the CBS peak disappears. In such a situation, one has to use directly $\ccfs$ to characterize the transition.

 The time evolution of $\Lambda$ in the three phases is shown in the right panel of Fig.~\ref{3DMomentumDistribution}. In the metallic phase $W\!<\!W_c$, a small CFS peak appears after a few $\tau$ and rapidly dies off, $\Lambda(t\!\to\!\infty)\to 0$. In the insulating phase $W\!>\!W_c$, the CFS peak steadily grows on the much longer Heisenberg time scale (see below) and eventually saturates to the CBS peak value, $\Lambda(t\!\to\!\infty)\to 1$. At the critical point, the peak very quickly saturates at $\Lambda(W_c)\!=\!\Lambda_c\!\approx\!  0.34$. In other words, localization triggers a macroscopic CFS peak, which is \textit{discontinuous} across the MIT, a behavior markedly different from the one of CBS.

\begin{figure}[h]
 \includegraphics[width=1\columnwidth]{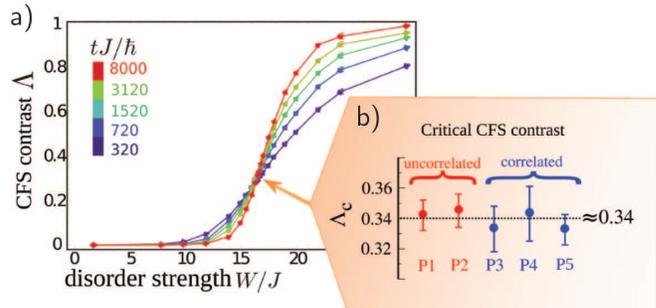}
 \caption{
(a) Normalized contrast $\Lambda$ as a function of disorder strength $W$ for different propagation times (at energy $E\!=\!J$). As time grows, $\Lambda$ converges toward a step function. All curves cross at $W\approx16.5J$, marking the critical point $W_c$ of the Anderson MIT. The corresponding contrast $\Lambda_c \approx 0.34$ is time-independent.
(b) Universality of the CFS contrast at the critical point. Uncorrelated disorder (red points): onsite box distribution (P1) and onsite Gaussian distribution (P2). Correlated disorder (blue points): onsite Gaussian distribution with Gaussian correlation (P3), blue detuned speckle (P4) and red-detuned speckle (P5). Error bars are discussed in the main text.}
\label{ContrastRatio}
\end{figure}

Fig. \ref{ContrastRatio}a shows $\Lambda$ as a function of $W$ for increasing times. The observed step is steeper as time increases, as expected for the behavior of a critical quantity across a phase transition where time plays the role of the system size. At long times, 
$\Lambda$ is 0 in the metallic regime and jumps to 1 in the insulating regime, irrespective of the exact value of $W$ chosen in each regime. Noticeably all curves cross at the critical point of the MIT, $W_c/J\!\approx\!16.5$, where $\Lambda_c\approx 0.34$. This important result reveals that exactly at the critical point, $\Lambda_c$ is time independent. As will be seen below, this value is {\it universal} and related to the multifractal properties of eigenstates at criticality.

\begin{figure}[ht]
 \includegraphics[width=0.76\columnwidth]{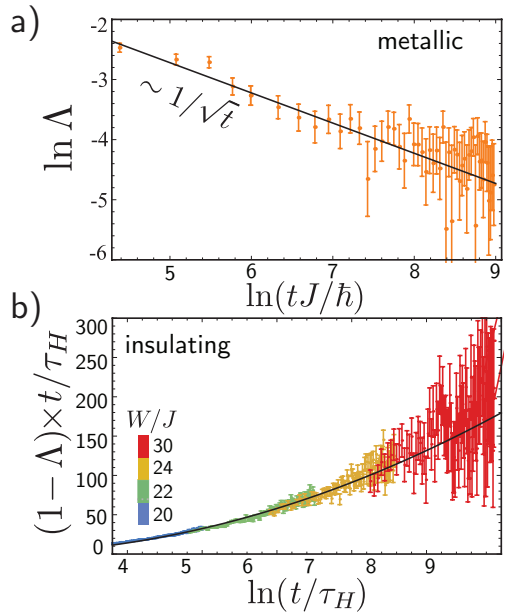}
 \caption{
 (Color online) Time dependence of the normalized contrast $\Lambda$. (a) in the metallic phase, the numerical points are well fitted by $1/\sqrt{t}$ (solid curve), in agreement with the theoretical prediction Eq. (\ref{CFSContrastDiffusive}). (b) in the insulating phase, the theoretical prediction Eq. (\ref{CFSContrastLocalized}), obtained in the limit $t\gg\tauh$, is confirmed with $\alpha=3.72$ and $\eta=0.129$ (solid curve). Here the Heisenberg time $\tauh$ has been computed independently by using the dynamics of the CBS peak width \cite{Ghosh15}.
 }
\label{CFSDynamics}
\end{figure}

In the metallic phase, perturbative techniques explain the long-time dynamics of CFS by a sum of two interference contributions, one featuring a concatenation of two  maximally-crossed diagram series and the other being its time-reversed counterpart \cite{Karpiuk12, Ghosh14, Micklitz14}. We find:
\begin{eqnarray}
W < W_c: \quad \Lambda(t) \ \sim\ \frac{1}{ 2\pi \hbar \rho D \sqrt{D t}}
\label{CFSContrastDiffusive}
\end{eqnarray}
where $\rho$ is the disorder-averaged density of states per unit volume (DOS) and $D$ the diffusion coefficient. 
Fig. \ref{CFSDynamics}a confirms that $\Lambda(t)$ behaves indeed as $1/\sqrt{t}$ in the metallic phase. In the insulating phase, 
we expand the initial state $|\bk_0\rangle$ on the localized eigenstates $|\varphi_n\rangle$ (with energy $\epsilon_n$) of $\hat{H}$ and get:
\begin{equation}
n(\bk,t) = \av{\big|\sum_{n}{\varphi_n^*(\bk_0)\varphi_n(\bk) \,\rme^{-\rmi \epsilon_n t/\hbar}\,}\big|^2},
\label{Eq:non-diagonal}
\end{equation}
where the sum only includes states with energies $\epsilon_n$ close to $E$.
In the long time limit, the off-diagonal oscillatory terms in the square wash out to 0, so that:
\begin{equation}
\label{diag_approx}
 n(\bk,t\to\infty)=\av{\sum_n{|\varphi_n(\bk_0)|^2|\varphi_n(\bk)|^2}}.
\end{equation}
Since our system has the TRS, its localized eigenstates can be chosen real in space and $\varphi_n(-\bk)=\varphi_n^*(\bk)$. Thus $n(-\bk,t\to\infty)=n(\bk,t\to\infty)$ and CFS and CBS become exact twin peaks in the long time limit. When the energy is fixed (like in our simulations), their same contrast is $C_\infty=\av{|\varphi_n(\bk_0)|^4}/(\av{|\varphi_n(\bk_0)|^2})^2-1$. 
This value is governed by the statistics of the $\varphi_n$~\cite{Ghosh14, Lee14}. When the localization length $\xi_\mathrm{loc}$ is much larger than the lattice spacing, the Hamiltonian inside a localization volume can be described by random matrix theory (RMT) in the GOE class. The $\varphi_n(\bk_0)$ are then independent random complex Gaussian variables and $C_\infty=1$. This leads to $\ccfs(t\to\infty)=\ccbs(t\to\infty)=C_\infty=1$, so that) $\Lambda=1$, in agreement with our numerical observations.  
It is only in the deeply localized regime, where the localization length becomes comparable to the lattice spacing, that $C_\infty$ decays slightly below unity. This leaves nevertheless $\Lambda=1$.
In the diffusive regime and in the vicinity of the critical point (on both sides), one also has $\ccbs(t\to\infty)=C_\infty=1$, so that the critical value $\Lambda_c\approx 0.34$ reflects entirely the behavior
of the CFS contrast.

The behavior of $\Lambda$ at long --but not infinite-- times can also be computed from Eq. (\ref{Eq:non-diagonal}). Indeed, within RMT the $\varphi_n(\bk)$ and the  $\epsilon_n$ are statistically uncorrelated variables and the average of each term in the expansion of the square in Eq.~(\ref{Eq:non-diagonal}) breaks into product of averages over the $\varphi_n(\bk)$ and the $\epsilon_n$. The latter is proportional to the Fourier transform of the DOS-DOS correlator $K(\omega)=\overline{\rho(E+\hbar\omega/2)\rho(E-\hbar\omega/2)}/\rho^2-1$, i.e. to the spectral form factor $K(t)$, leading to $\Lambda(t)=2\pi\hbar\rho M^3 K(t)$~\cite{Lee14, Ghosh14}. Following the correlated volume approach~\cite{Mott70}, we obtain $K(t)$ by estimating the 3D hybridization of localized states with energies lying within a mean level spacing $\Delta = 2\pi\hbar/\tauh$. 
This gives 
$K(\omega) \sim \delta(\hbar\rho M^3\omega) + (\xi_\mathrm{loc}/M)^3\ \ln^3(|\omega|\tauh/4\pi)$ for $|\omega|\tauh\ll1$.
After Fourier transform, we obtain
\begin{eqnarray}
W>W_c: \quad \Lambda(W,t)  \approx 1-\alpha \frac{\ln^2( \eta t/\tauh )}{(t/\tauh)}.
\label{CFSContrastLocalized}
\end{eqnarray}
The phenomenological constants $\alpha$ and $\eta$ respectively account for subdominant corrections in the distribution of localized states and for a possible numerical prefactor in the definition of $\tauh$. The time scale $\tauh=2\pi\hbar\rho\xi_\mathrm{loc}^3= 2\pi\hbar/\Delta$ is the Heisenberg time, i.e. the inverse of the mean level spacing $\Delta$ within a localization volume. It is the typical time beyond which off-diagonal terms in Eq. (\ref{Eq:non-diagonal}) average to zero. Note that the previous reasoning is invalid in the metallic phase since eigenstates are delocalized over an infinite volume: no minimum energy scale can show up in the expansion Eq.~(\ref{Eq:non-diagonal}) and off-diagonal terms never average to zero.

\begin{figure}[ht]
 \includegraphics[width=1\columnwidth]{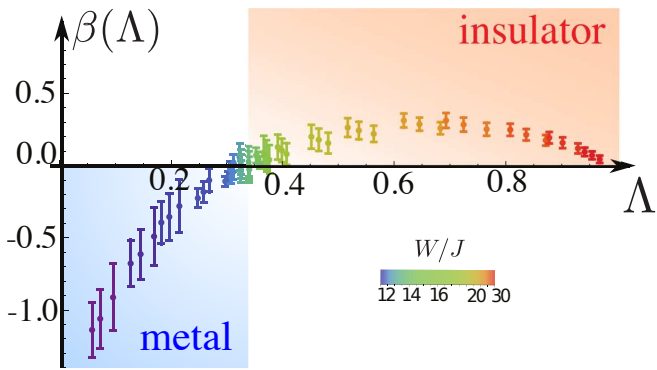}
 \caption{(Color online) CFS scaling function $\beta(\Lambda)$ obtained by compiling data for $\Lambda$. 
Each color corresponds to a given value of $W$ and, for each $W$, each point corresponds to a different propagation time $t$. All points fall on a single curve. The function changes sign at the critical point where $\Lambda\!=\!\Lambda_c\!\simeq\! 0.34$. 
Eqs.~(\ref{CFSContrastDiffusive}) and (\ref{CFSContrastLocalized}) give the asymptotic
	behaviors $\beta(\Lambda\!\to\!0)\!=\!-3/2$ in the metallic phase and 
	$\beta\!\approx\!3(1-\Lambda)$ in the insulating phase.}
 \label{BetaFunction}
\end{figure}

As shown by Eq. (\ref{CFSContrastLocalized}), $\Lambda$ depends on $W$ and $t$ only through the parameter $t/\tauh$. This property is confirmed numerically in Fig. \ref{CFSDynamics}b, where all numerical points obtained for different $W$ collapse onto a single curve. 
 This one-parameter scaling law can be extended to the whole range of disorder strengths if one introduces a ``system size'' $L=[t/2\pi\hbar\rho]^{1/3}$ and defines a correlation length $\xi \propto \xi_\mathrm{loc}$ in the insulating phase and $\xi \propto (2\pi\hbar\rho D)^{-1}$ in the metallic one \cite{Ghosh14}. Then, both Eqs.~(\ref{CFSContrastDiffusive}) and (\ref{CFSContrastLocalized}) depend on $L/\xi$ only, suggesting that  $\Lambda$ is a natural one-parameter scaling observable to study an Anderson MIT. Following the historical scaling theory of AL \cite{Abrahams79}, the scaling function $\beta=d(\ln \Lambda)/d(\ln L)$ should depend on $\Lambda$ only. This is confirmed in Fig.~\ref{BetaFunction}, where points obtained by computing numerically $\Lambda$ at various times and disorder strengths fall all on the same curve. As time increases, the system goes metallic when $\beta(\Lambda)<0$ and insulating when $\beta(\Lambda)>0$. The critical phase is signaled by the fixed point $\beta(\Lambda_c)=0$. 
 In the vicinity of the MIT, the correlation length diverges as $\xi\!\propto\!|W-W_c|^{-\nu}$, where $\nu$ is the critical exponent. To accurately determine the critical parameters $W_c$ and $\nu$, we use a finite-time scaling analysis which consists in writing $\Lambda= F(\chi L^{1/\nu})$ with $|\chi|\!\propto\!\xi^{-1/\nu}$ (scaling hypothesis) 
and fitting the numerical data with a double Taylor expansion $\Lambda=\sum_{n=0}^{n_R} F_n \chi^nL^{n/\nu}$ and $\chi=\sum_{m=1}^{m_R}b_m(W-W_c)^m$ where $F_n$, $b_m$, $W_c$ and $\nu$ are the fit parameters \cite{Ghosh15}. With our 1095 data points, we obtain a good fit for $n_R=m_R=2$, achieving a chi-square per degree of freedom $\approx1.6.$ The uncertainties of $W_c$ and $\nu$ are obtained by dividing the whole configuration sample into several independent subsets and estimating  $W_c$ and $\nu$  for each subset. This approach gives $W_c/J\!=\!16.53\pm0.03$ and $\nu=1.51\!\pm\!0.07$. The result agrees very well with an independent numerical calculation using the transfer-matrix method~\cite{Kramer93, Slevin14}. 

An intriguing question is the physical meaning of $\Lambda_c\equiv \Lambda(W_c)$. In Fig.~\ref{ContrastRatio}b we show $\Lambda_c$ for different models of disorder potentials: (P1) the uncorrelated box distribution used throughout the paper; (P2) uncorrelated disorder with Gaussian on-site distribution $P(V_i)\propto \exp(-V_i^2/W^2)$; (P3) Gaussian on-site distribution, with spatial Gaussian correlation $C(\br)=W^2\exp(-\br^2/\zeta^2)$ where $\zeta$ is the correlation length; (P4) blue-detuned speckle potential, $P(V_i)\propto\exp(-V_i/W)$ for $V_i>0$ and $C(\br)=W^2 [\sin(|\br|/\zeta)/(|\br|/\zeta)]^2]$; (P5) red-detuned speckle, obtained from (P4) by $V_i\to -V_i$. 
To obtain $\Lambda_c$, we first locate the mobility edges $W_c$ for each disorder model by using the transfer-matrix method, and then compute the normalized CFS contrast $\Lambda_c$ from the propagation of a plane wave at $W=W_c$. Error bars on $\Lambda_c$ include both the finite accuracy in the estimation of  $W_c$ and the statistical error in the determination of $\Lambda_c$. 
The validity of this approach is confirmed by a finite-size scaling analysis (see above) of model (P1), yielding an independent estimate $\Lambda_c\!=\!0.329\pm 0.015$ compatible with the one in Fig.~\ref{ContrastRatio}b. Fig.~\ref{ContrastRatio}b demonstrates that $\Lambda_c$ is insensitive to the microscopic details of the disorder and thus \textit{universal}: different on-site energy distributions and spatial correlations lead all to the same $\Lambda_c$ while the critical disorder $W_c$ varies strongly. 

Assuming that the relation to the form factor still holds at the critical point, we infer 
$\Lambda_c=\lim_{t\to 0} 2\pi\hbar\rho M^3 K(t)=\kappa(W_c)\equiv \kappa_c$, a positive quantity quantifying the statistical fluctuations of the energy spectrum known as the \textit{spectral compressibility} \cite{Chalker96}. In the metallic phase, the spectrum is rigid -- approximately described by GOE random matrices -- and fluctuations are small: $\kappa \to 0$. In the insulating phase, fluctuations are large and $\kappa=1$. At the mobility edge, $\kappa_c$ takes on an intermediate value depending only on the universality class of the MIT and carrying information on the multifractal character of the critical eigenstates \cite{Zharekeshev95}. It has been conjectured that  $\kappa_c\!=\!1-D_1/3$, where the ``information dimension'' $D_1$ gives the amount of entropy of the critical eigenstates \cite{Bogomolny11}.
For the Anderson model, $D_1\!=\!1.958\pm 0.005$~\cite{Rodriguez11}, which leads to the prediction $\Lambda_c\!=\!\kappa_c\!=\!0.347,$ in excellent agreement
with the numerically measured value $\Lambda_c\!=\!0.342\pm0.01$. The alternate conjecture~\cite{Chalker96b} $2\kappa_c\!=\!1-D_2/3$ predicts $\kappa_c\approx 0.29$, deviating significantly from our numerical results. This demonstrates that the CFS peak at criticality is a direct universal experimental probe of $D_1$, independent of the disorder distribution and spatial correlation. 

In conclusion, we have shown that CFS constitutes an experimentally measurable order parameter for Anderson MITs. The peak contrast obeys a one-parameter scaling law, gives direct access to properties which are in general extremely difficult to assess, such as the critical exponents, and exhibits a universal value at criticality related to the multifractal properties of eigenstates. Unlike CBS, which is absent when TRS is broken, CFS is robust, universal and does not require any specific symmetry. It could thus be used to characterize different types of MITs beyond the conventional GOE class.

SG acknowledges the support of the PHC Merlion Programme of the French Embassy in Singapore. This work was granted access to the HPC resources of TGCC under the allocations 2015-057083 and 2016-057644 made by GENCI (Grand Equipement National de Calcul Intensif) and to the HPC resources of MesoPSL financed by the Region Ile de France and the project Equip@Meso (reference ANR-10-EQPX-29-01) of the programme Investissements d'Avenir supervised by the Agence Nationale pour la Recherche. This research is supported by the National Research Foundation, Prime Minister's Office, Singapore and the Ministry of Education, Singapore under the Research Centres of Excellence programme.


\end{document}